# Remote smartphone-based speech collection: acceptance and barriers in individuals with major depressive disorder

*Judith Dineley[1], Grace Lavelle[2], Daniel Leightley[2], Faith Matcham[2], Sara Siddi[3],
Maria Teresa Peñarrubia-María[4], Katie M. White[2], Alina Ivan[2], Carolin Oetzmann[2],
Sara Simblett[2], Erin Dawe-Lane[2], Stuart Bruce[5], Daniel Stahl[2], Yatharth Ranjan[2],
Zulqarnain Rashid[2], Pauline Conde[2], Amos A. Folarin[2,6], Josep Maria Haro[3], Til Wykes[2,7],
Richard J.B. Dobson[2,6], Vaibhav A. Narayan[8], Matthew Hotopf[2,7], Björn W. Schuller[1,9],
Nicholas Cummins[1,2], The RADAR-CNS Consortium[10]*

[1]Chair of Embedded Intelligence for Health Care and Wellbeing, University of Augsburg, Germany
[2] Institute of Psychiatry, Psychology and Neuroscience, King's College London, London, UK
[3]Parc Sanitari Sant Joan de Déu, Fundació Sant Joan de Déu, CIBERSAM, Barcelona, Spain
[4]Centro de Salud Bartomeu Fabrés Anglada, Gavà, IDIAP Jordi Gol, Spain
[5]RADAR-CNS Patient Advisory Board, King's College London, UK
[6]Institute of Health Informatics, University College London, London, UK
[7]South London and Maudsley National Health Services Foundation Trust, London, UK
[8]Janssen Research and Development LLC, Titusville, NJ, United States
[9]GLAM - Group on Language, Audio, & Music, Imperial College London, London, UK
[10]www.radar-cns.org

`judith-anne.dineley@uni-a.de, nick.cummins@kcl.ac.uk`

## Abstract

The ease of in-the-wild speech recording using smartphones has sparked considerable interest in the combined application of speech, remote measurement technology (RMT) and advanced analytics as a research and healthcare tool. For this to be realised, the acceptability of remote speech collection to the user must be established, in addition to feasibility from an analytical perspective. To understand the acceptance, facilitators, and barriers of smartphone-based speech recording, we invited 384 individuals with major depressive disorder (MDD) from the Remote Assessment of Disease and Relapse – Central Nervous System (RADAR-CNS) research programme in Spain and the UK to complete a survey on their experiences recording their speech. In this analysis, we demonstrate that study participants were more comfortable completing a scripted speech task than a free speech task. For both speech tasks, we found depression severity and country to be significant predictors of comfort. Not seeing smartphone notifications of the scheduled speech tasks, low mood and forgetfulness were the most commonly reported obstacles to providing speech recordings.

## 1. Introduction

Speech, both as a natural language processing (NLP) and computational paralinguistics signal, is increasingly regarded as a key digital phenotype for a range of health conditions [1], [2], particularly mental and neurological disorders [3], [4]. With the increasing ubiquity of remote measurement technology (RMT), such as smartphones, wearables and other Internet-of-Things (IoT) devices, how we collect speech in speech-health studies is also changing.

In-the-wild speech collection [5]–[9] enables research study participants to incorporate speech recordings into their daily lives, providing them with the flexibility of when and where to record their speech. Variables that may influence the paralinguistic and linguistic content of someone's speech, such as depression severity, can also be measured at the same time. However, moving from the laboratory to real-life environments comes with the need to understand participants' motivations, facilitators and barriers that may affect the remotely-collected speech data.

There have been minimal efforts to understand the challenges of remote speech collection. Simblett et al. [10], conducted a systematic review of RMT approaches as a healthcare management tool and identified that most studies did not assess the usability and acceptability of RMT. The studies reviewed also did not target any specific RMT signal, so no speech-specific conclusions can be drawn from this literature.

Few studies in individuals with mental health disorders have explicitly explored user engagement with in-the-wild speech collection. One study [11] collected speech samples to track clinical changes in 47 individuals with serious mental illnesses. The authors implemented an exit survey, where 22 of 24 respondents gave positive statements on their experience. While promising, the survey was not fully described, and only limited results were reported. The study also collected speech via a dial-in system accessible using any phone, so participants did not actually interact with any RMT.

Given the growing number of speech-health studies [12]–[21], understanding the factors related to motivation, convenience, accessibility and usability of remote speech data collection tools is vital to maximise the research value of this unique health signal. This paper describes a survey designed to understand some of these factors in individuals diagnosed with major depressive disorder (MDD) and enrolled in the multi-centre *Remote Assessment of Disease and Relapse – Central*

*Nervous System* (RADAR-CNS) research programme [22], an observational, longitudinal, prospective study collecting multimodal RMT data in six European countries.

Based on survey responses, we specifically report how comfortable participants felt completing scripted and free speech tasks, and differences in comfort level with sociodemographic factors and depression severity. We also quantify the obstacles perceived by participants to have prevented speech collection, and the accuracy of participants' reporting of their level of speech task completion.

## 2. Methodology

### 2.1 Speech collection in RADAR-CNS

Participants with MDD were asked to complete two speech-recording tasks every two weeks at the same time as questionnaires assessing depression severity [23] and self-esteem [24]. In a scripted speech task (SS), the participants read aloud an extract from Aesop's fable, *The North Wind and the Sun* [25]. The second task was a free-speech activity (FS) in which participants were asked to speak about what they were looking forward to in the following seven days [3], [26].

Speech collection began in the UK in August 2019 and in December 2019 in Spain. Recordings were collected using the open-source RADAR-base m-health platform [27]. Participants recorded their speech on smartphones using a custom-designed app that produced notifications each time speech recordings were scheduled. English and Castilian Spanish versions of the app were available. Once recorded, speech data were encrypted and sent to a secure server.

### 2.2 Other data collected

Sociodemographic data, including age at enrolment, gender and years in education, were collected for all participants. Depression severity was also assessed using the Inventory of Depressive Symptomatology – Self-Reported (IDS-SR) [28], conducted every three months as part of the RADAR-CNS MDD protocol [29]. Participants' individual speech task completion rates were calculated as a percentage of the scheduled tasks.

### 2.3 Speech survey design & implementation

The survey was designed to minimise the burden on study participants, who already had several tasks to complete as part of RADAR-CNS. It was developed in a smartphone-compatible format using the Qualtrics survey software platform, which predicted a completion time of eight minutes. Participants were asked a maximum of 12 questions that were mostly multiple choice, all of which were optional (see associated multimedia file). Questions were developed based on anecdotal feedback from participants, discussions with service users in the RADAR-CNS Patient Advisory Board and an RMT engagement survey reported in the literature [30].

The first question asked participants how often they completed the speech recordings on a five-point scale from *every time* to *never*. The next two questions asked how comfortable participants felt completing each speech task, also on a 5-point scale, from *extremely uncomfortable* to *extremely comfortable*. Two further questions asked participants to select the barriers that prevented them from completing speech recordings at least once, divided loosely into *practical* and *other,* with a total of 17 possible barriers. Respondents could also name barriers not listed. These questions addressed the issues of perceived speech task completion, acceptance and barriers to completion; these were identified as core factors to examine in this study.

Participants received survey invitations on 14th October 2020 (UK) and on 4th November 2020 (Spain) via an emailed link. All MDD participants in the UK and Spain at that time were invited. The survey links were tagged with participants' unique, anonymous study identification numbers to extract matching sociodemographic, depression severity and speech completion data. Responses collected up to 25th January 2021 (UK) and 18th January 2021 (Spain) were pooled for analysis.

### 2.4 Survey analysis

**Response cohort:** The effect of sociodemographic variables (age, gender, years in education and country), depression severity and speech task completion rates in survey responders and non-responders were assessed using logistic regression with bootstrapping to assess the generalisability of the responses. Depression severity was quantified as each individual's IDS score within six weeks of the survey launch.

**Assessment of comfort:** Linear regression with bootstrapping was conducted to identify significant associations between comfort levels for both speech tasks and sociodemographic factors as well as depression severity. Analyses were undertaken using the IBM SPSS Statistics 27 software.

**Perceived barriers preventing speech recording:** The most commonly reported barriers were identified, in addition to the range, mean and standard deviation of the number of barriers.

**Accuracy of self-reported speech task completion:** Self-reported completion was compared with actual speech task completion obtained via the participant identification numbers.

A member of the RADAR-CNS Patient Advisory Board provided feedback on the results of these analyses.

## 3. Results and Discussion

### 3.1 Response cohort

Survey links were emailed to the 384 participants (UK: 254, Spain: 130), who comprised 76% of the total of 505 enrolments in the two countries. We received responses from 209 participants (UK: 135, Spain: 74), a response rate of 54%. Table 1 details the sociodemographic data, depression severity and speech task completion rates of those who were invited to do the survey and those who responded.

Results of a logistic regression model with survey response status as the outcome and sociodemographic factors and depression severity as predictors indicate that older participants were significantly more likely to respond to the survey (Wald $\chi^2 = 8.80$, $p < 0.005$, odds ratio = 1.03). Gender (Wald $\chi^2 = 0.84$, $p > 0.05$, odds ratio = 1.31), years in education (Wald $\chi^2 = 0.004$, $p > 0.05$, odds ratio = 1.00), country (Wald $\chi^2 = 1.24$, $p > 0.05$, odds ratio = 0.70) and depression severity (Wald $\chi^2 = 0.45$, $p > 0.05$, odds ratio = 1.01), were all non-significant.

Participants who completed the speech survey demonstrated higher completion rates for the scheduled speech tasks than those who did not complete the survey. The average completion rates for survey responders were 51% (SS) and 44% (FS) versus 31% (SS) and 26% (FS) for non-responders. A logistic regression with completion rates of both speech tasks as predictor variables and survey response status as the outcome indicates that participants with higher SS completion were more likely to complete the survey (Wald $\chi^2 = 6.59$, $p = 0.01$, odds ratio = 6.93), though there was no association with FS completion (Wald $\chi^2 = 0.33$, $p > 0.05$, odds ratio = 1.58).

Table 1: *Sociodemographic characteristics and completion rates of scripted speech (SS) and free speech (FS) of those invited to do the speech survey and those who responded.*

| n | | Invited<br>384 | Responded<br>209 |
|---|---|---|---|
| gender | female | 284 | 149 |
| | male | 97 | 59 |
| Age at enrolment (years) | mean | 48 | 51 |
| | range | 18-80 | 18-80 |
| | SD | 14.5 | 13.4 |
| IDS score | mean | 31 (moderate) | 32 (moderate) |
| | range | 0-74 | 2-74 |
| | SD | 15.5 | 15.4 |
| Years of education | mean | 15 | 15 |
| | range | 4-26 | 5-26 |
| | SD | 4.1 | 4.1 |
| SS completion (%) | mean | 42 | 51 |
| | SD | 31 | 30 |
| | range | 0-100 | 0-100 |
| FS completion (%) | mean | 36 | 44 |
| | SD | 29 | 30 |
| | range | 0-100 | 0-100 |

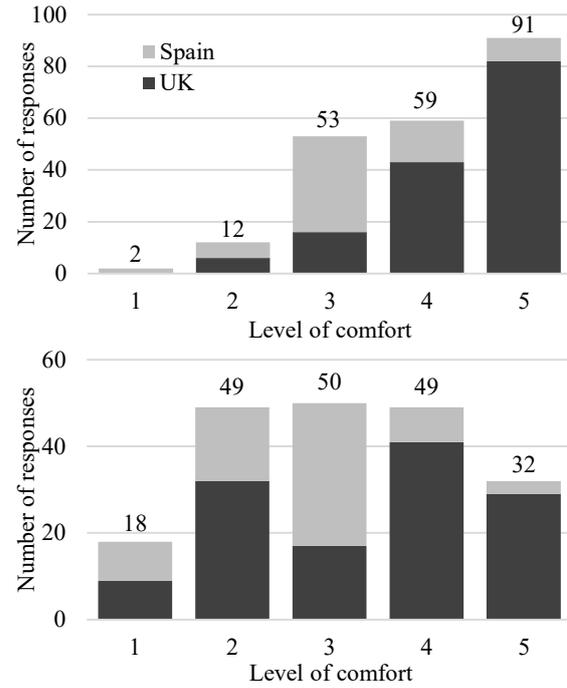

Figure 1: *Self-reported comfort level recording scripted (top) and free speech (bottom) on a 5-point scale from 1, 'extremely uncomfortable', to 5, 'extremely comfortable'.*

Table 2: *Linear regression of participant-reported comfort in recording scripted speech (SS) and free speech (FS) vs IDS score and sociodemographic variables. bs: standardised b.*

| Variable | Scripted Speech (SS) | | | | Free Speech (FS) | | | |
|---|---|---|---|---|---|---|---|---|
| | bs | b | *p* | 95% CI | bs | b | *p* | 95% CI |
| IDS score | -.191 | -.013 | *.009* | -.022, -.004 | -.282 | -.023 | *.001* | -.033, -.011 |
| Country (UK = 1, Spain = 2) | -.452 | -.958 | *.001* | -1.308, -.618 | -.243 | -.616 | *.003* | -1.026, -.202 |
| Age | .035 | .003 | .622 | -.007, .014 | 0.006 | .001 | .952 | -.015, .016 |
| Gender (male = 0, female = 1) | -.034 | -.074 | .562 | -.359, .209 | -.091 | -.240 | .203 | -.625, .205 |
| Years of education | -.006 | -.001 | .941 | -.0423, .039 | -.119 | -.036 | .142 | -.080, .014 |

This result is unsurprising; the SS task appears before the FS task in the study app, so participants who did not complete the SS task are less likely to have completed the FS task. Also, as the survey indicates (Section 3.2), participants are less comfortable completing the free speech task.

### 3.2 Comfort levels in recording speech

The most frequently reported comfort levels were *extremely comfortable* (SS) and *neither comfortable nor uncomfortable* (FS) (Figure 1). Responses to the FS task are likely to contain personal information, and so the task may provoke an emotional response [3], [26]. This could have made participants more uncomfortable than when reading aloud a set text. Restrictions due to the COVID-19 pandemic could also have made thinking of things to say in the FS task more challenging.

Linear regression demonstrated that depression severity and country were significant predictors of comfort in both speech tasks (Table 2, Figure 2). We speculate that the higher cognitive impairment and fatigue associated with severe depression [31], [32] contributes to the decreased comfort seen with increasing depression severity in both tasks.

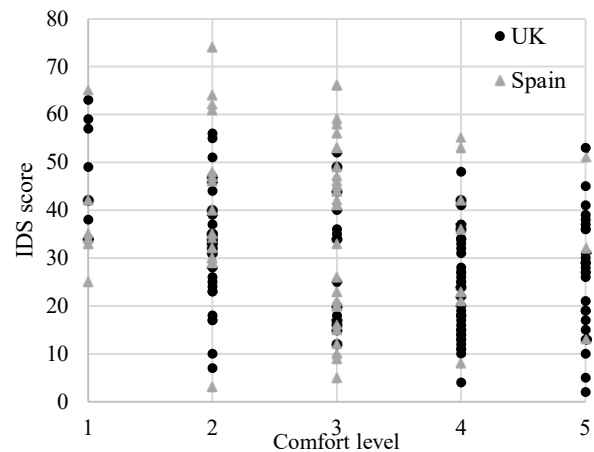

Figure 2: *A scatter plot of depression severity, quantified by IDS scores, versus participant comfort in recording free speech. Comfort is quantified on a 5-point scale from 1, 'extremely uncomfortable', to 5, 'extremely comfortable'.*

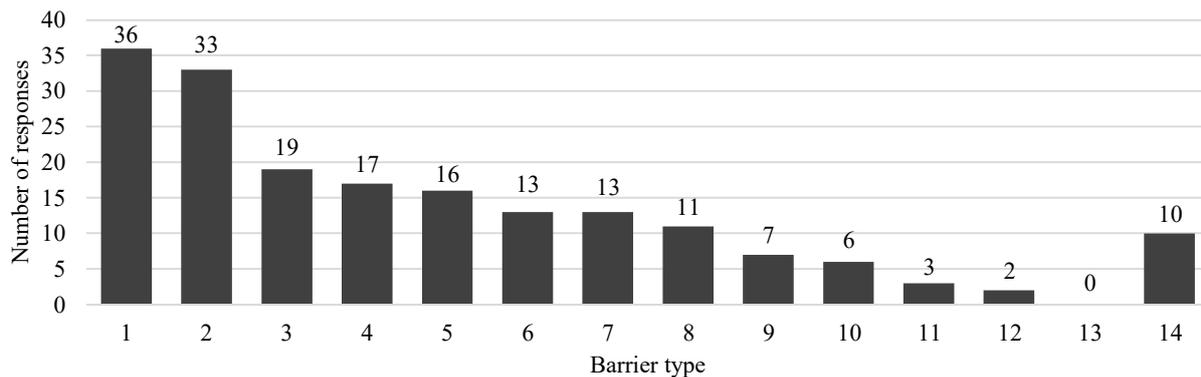

Figure 3: *Self-reported obstacles preventing speech task completion. 1 – didn't see the notifications, 2 – low mood, 3 – forgot 4 – felt unwell in some other way or too tired, 5 – couldn't find a quiet place, 6 – felt self-conscious, 7 – technical problems with phone or app, 8 – didn't have time, 9 – already contribute enough to RADAR-CNS, 10 – privacy concerns, 11 – not interested in recording speech, 12 – speech part of the app difficult to use, 13 – instructions not clear, 14 – other.*

Depression severity was a stronger predictor of comfort than country in SS, while country was a stronger predictor of comfort in FS. Age, gender and years of education were not significant predictors of comfort in either speech task. The $R^2$ values (SS: 0.29, FS: 0.16), suggest other factors could also account for the difference in comfort levels.

### 3.3 Perceived barriers in recording speech

Barriers that prevented participants from completing speech recordings on at least one occasion included practical and technical barriers, such as being unable to find a quiet place to record, and other issues such as low mood (Figure 3). In total, 104 (50%) of survey responders reported encountering at least one such barrier. The number of barriers reported per participant ranged from zero to eight (mean = 0.9, SD = 1.2).

The most commonly reported barriers were (1) not seeing the app notifications for the speech tasks (36 responses), (2) low mood (33 responses) and (3) forgetfulness (19 responses). RADAR participants receive up to four notifications at the same time – for depression, self-esteem and COVID-19 questionnaires, followed by speech. Potentially, after completing the questionnaires, participants are less motivated to spend time and effort on the speech task, and this could be exacerbated with more severe depression.

### 3.4 Self-reported completion versus true completion

The majority of participants responding to the survey reported completing the speech task every time or on most occasions that it was scheduled. However, in a striking finding, many respondents markedly overestimated their completion rates in both speech tasks (Figure 4). We speculate that several factors contribute to this discrepancy. For example, after completing depression and self-esteem questionnaires, plus a COVID-19 questionnaire since the pandemic began, participants may not notice the notifications for the speech task. Other factors may include recall and social desirability biases, as well as missed app notifications.

## 4. Conclusions

As the popularity of *in-the-wild* data collection paradigms grow in the speech community, we need to better understand the associated barriers and facilitators, particularly in health applications. To the best of the authors' knowledge, this preliminary analysis is the first to focus on the experiences of individuals recording their speech using RMT for health research. Gaining insights in this area is vital to maximise research value and participant comfort in future studies. The core issues described raise several questions for investigation that will be addressed in future work. These include the investigation of technical issues that lower speech recording rates and scenarios that may cause participants to overestimate their speech recording rates. The RADAR-CNS speech survey has since been released at the remaining MDD study site and all three sites in the multiple sclerosis arm of the study.

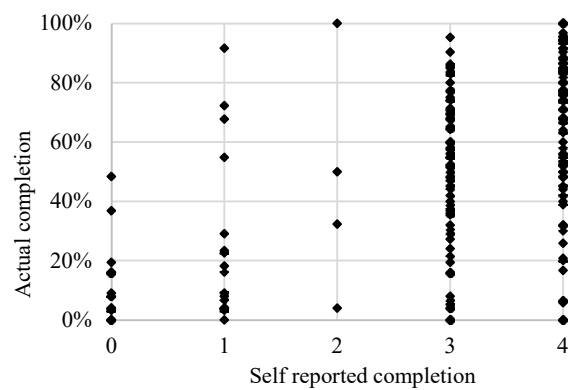

Figure 4: *Actual completion vs self-reported completion, FS task. Self-reported completion is on a 5-point scale from 0 (Never) to 4 (Every time – every 2 weeks).*

## 5. Acknowledgements

The RADAR-CNS project has received funding from the Innovative Medicines Initiative 2 Joint Undertaking under grant agreement No 115902. This Joint Undertaking receives support from the European Union's Horizon 2020 research and innovation programme and EFPIA (www.imi.europa.eu). This communication reflects the views of the RADAR-CNS consortium and neither IMI nor the European Union and EFPIA are liable for any use that may be made of the information contained herein. The funding bodies have not been involved in the design of the study, the collection or analysis of data, or the interpretation of data.

We would like to thank all members of the RADAR-CNS patient advisory board, who all have experience of living with or supporting those who are living with depression, epilepsy or multiple sclerosis.